\documentclass[aps,prl,twocolumn,showpacs,superscriptaddress,floatfix,10pt]{revtex4-1}
\usepackage{framed}
\usepackage{graphicx}
\usepackage{amsmath}
\usepackage{lmodern}
\usepackage{epsfig}
\usepackage{helvet}
\usepackage{framed}
\usepackage{bbold}
\usepackage{subfigure}
\usepackage{hyperref}

\usepackage{bm}

\hypersetup{
  colorlinks   = true, 
  urlcolor     = blue, 
  linkcolor    = blue, 
  citecolor   = red, 
}

\newcommand{\ket}[1]{\mbox{$| #1 \rangle$}}

\allowdisplaybreaks[3]
\def\letter{paper } 
\def\appendix{Appendix }

\newcommand{\beq}{\begin{equation}}
\newcommand{\eeq}{\end{equation}}
\newcommand{\beqa}{\begin{eqnarray}} 
\newcommand{\eeqa}{\end{eqnarray}}
\newcommand{\blgn}{\begin{aligned}}
\newcommand{\elgn}{\end{aligned}}

\newcommand{\la}{\langle}
\newcommand{\ra}{\rangle}

\newcommand{\f}{\frac}

\newcommand{\pa}{\partial}

\allowdisplaybreaks[3]

\def\letter{paper } 

\begin{document}

\title{Continuous tensor network renormalization for quantum fields} 

\author{Q. Hu}
\email{qhu@perimeterinstitute.ca}
\affiliation{Perimeter Institute for Theoretical Physics, Waterloo, Ontario N2L 2Y5, Canada}  \date{\today}
\affiliation{Department of Physics and Astronomy, University of Waterloo, Waterloo, Ontario N2L 3G1,
Canada}
\author{A. Franco-Rubio}
\affiliation{Perimeter Institute for Theoretical Physics, Waterloo, Ontario N2L 2Y5, Canada}  \date{\today}
\affiliation{Department of Physics and Astronomy, University of Waterloo, Waterloo, Ontario N2L 3G1,
Canada}

\author{G. Vidal}
\affiliation{Perimeter Institute for Theoretical Physics, Waterloo, Ontario N2L 2Y5, Canada}  \date{\today}

\date{\today}

\begin{abstract}
On the lattice, a renormalization group (RG) flow for two-dimensional partition functions expressed as a tensor network can be obtained using the tensor network renormalization (TNR) algorithm [G. Evenbly, G. Vidal, Phys. Rev. Lett. 115 (18), 180405 (2015)]. In this work we explain how to extend TNR to field theories in the continuum. First, a short-distance length scale $1/\Lambda$ is introduced in the continuum partition function by smearing the fields. The resulting object is still defined in the continuum but has no fluctuations at distances shorter than $1/\Lambda$. An infinitesimal coarse-graining step is then generated by the combined action of a \textit{rescaling} operator $L$ and a \textit{disentangling} operator $K$ that implements a quasi-local field redefinition. As demonstrated for a free boson in two dimensions, continuous TNR exactly preserves translation and rotation symmetries and can generate a proper RG flow. Moreover, from a critical fixed point of this RG flow one can then extract the conformal data of the underlying conformal field theory.
\end{abstract}

\maketitle

The study of many-body systems is a major challenge of modern physics. Following the seminal work of Kadanoff \cite{Kadanoff} and Wilson \cite{Wilson}, the renormalization group (RG) allows us to investigate how the physics of a many-body system changes with scale while providing a conceptual framework for understanding universality in second order phase transitions. More recently, ideas from quantum information have led to a new generation of powerful numerical RG algorithms for lattice systems \cite{TRG,TRG2,TRG3,TRG4,TRG5,TRG6,TRG7,TNR,TNRyieldsMERA,TNRalgorithms, TNRlocal,loopTNR,skeleton,TNRplus,giltTNR}, starting with the breakthrough work of Levin and Nave \cite{TRG}, who wrote the partition function of a two-dimensional statistical model as a tensor network and proposed coarse-graining the latter by applying linear algebra compression methods 
(see \cite{TRG2,TRG3,TRG4,TRG5,TRG6,TRG7} for related algorithms). 
Building on that proposal, Evenbly and Vidal subsequently introduced the tensor network renormalization (TNR) algorithm \cite{TNR,TNRyieldsMERA,TNRalgorithms, TNRlocal}, which was shown to generate a proper RG flow $\left\{ A_1, A_2, \cdots \right\}$ in the space of tensors $A$ 
(see \cite{loopTNR, TNRplus, skeleton, giltTNR} for similar proposals). 
In particular, when applied to a critical lattice model, TNR naturally flows to an RG fixed point, thus explicitly realizing scale invariance on the lattice. From the fixed-point tensor $A_{\star}$ one can then extract the critical universality class of a second order phase transition, namely the conformal data that characterizes the underlying conformal field theory (CFT) \cite{CFT1,CFT2,CFT3}. TNR, which recently inspired research in the context of the holographic principle of quantum gravity \cite{TNRholo1,TNRholo2,TNRholo3,TNRholo4}, can be readily applied also to field theories by bringing them to the lattice. However, the discretization procedure destroys the field theory spacetime symmetries (continuous translation and rotation symmetries), which are only retained in an approximate, emergent sense. This makes it of interest, both conceptually and computationally, to extend TNR to field theories directly in the continuum, in such a way that the original spacetime symmetries are explicitly preserved. 

In this \letter we propose an extension of TNR to the continuum. As a preliminary step, fluctuations at distances shorter than a cut-off $1/\Lambda$ are removed from the Euclidean partition function by smearing the quantum field degrees of freedom. Then an infinitesimal transformation of continuous TNR (cTNR) is generated by the combined action of two operators --a \textit{rescaling} operator $L$ that rescales space and a \textit{disentangling} operator $K$ that implements a quasi-local field redefinition. When applied to the simple case of a free boson field theory, cTNR is seen to generate the correct RG flow, including a critical fixed point for the massless theory. As in lattice TNR, from this fixed point we can extract the conformal data of the underlying CFT. Our construction resembles, but is not equivalent to, the proposal of the continuous version of the multiscale entanglement renormalization ansatz (MERA) \cite{MERA1,MERA2}, known as continuous MERA (cMERA) \cite{cMERA}, for ground states of field theories, which is so far only well-understood for free theories but has nevertheless attracted much attention in the context of holography as a potential toy model realization of the AdS/CFT correspondence \cite{cMERAholo1,cMERAholo2,
cMERAholo3,cMERAholo4,cMERAholo5,cMERAholo6,cMERAholo7,
cMERAholo8,cMERAholo9}.

\begin{figure}[t]
\includegraphics[width=\columnwidth]{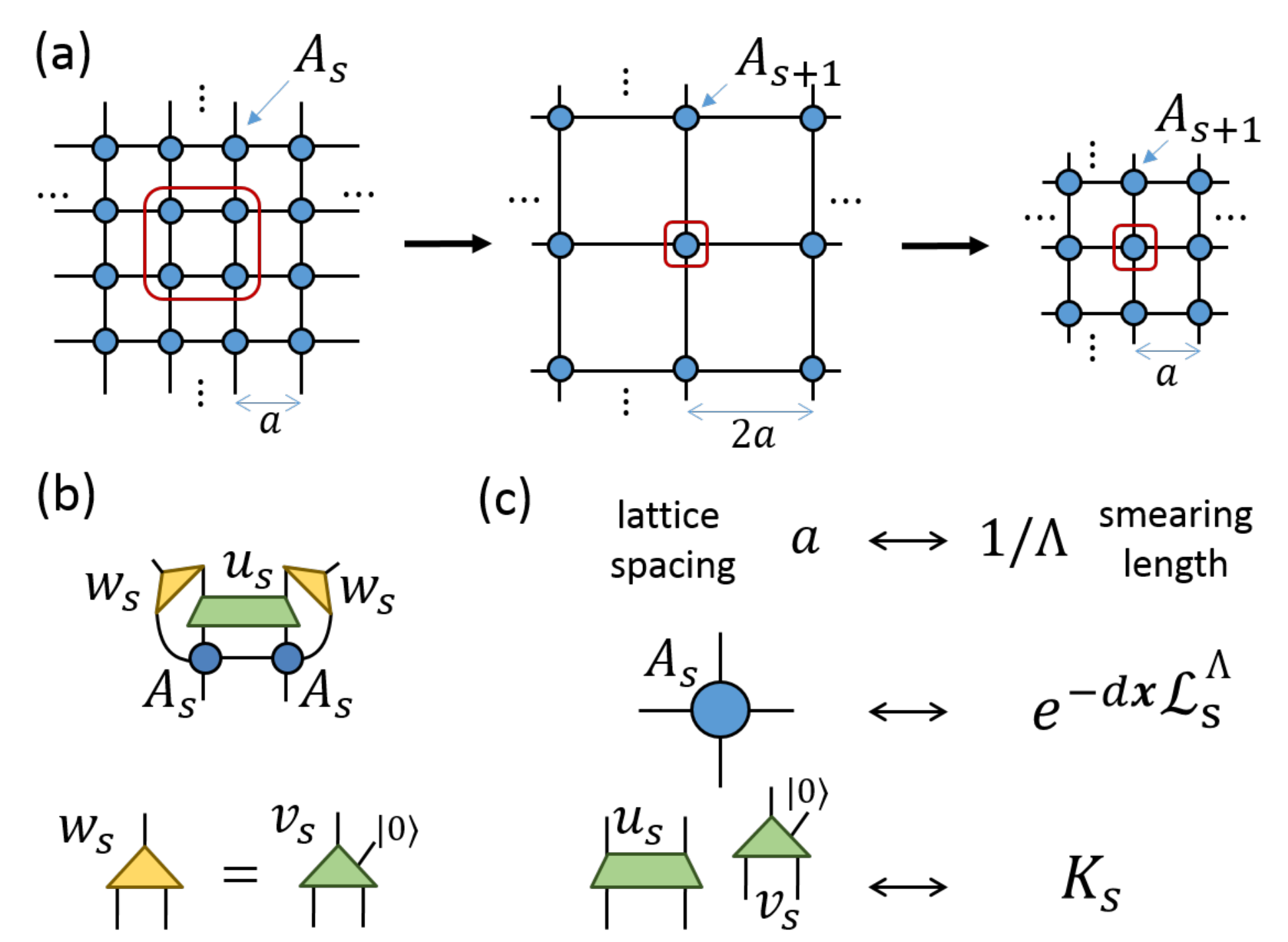}
\caption{
(a) TNR acts of a network of tensors $A_s$ representing a partition function on a lattice with spacing $a$, by replacing a block of four tensors $A_s$ with a single tensor $A_{s+1}$, then rescaling the lattice spacing of resulting network from $2a$ back to $a$.
(b) As part of the intricate local manipulations that coarse-grain the network, TNR applies disentanglers $u_s$ and isometries $w_s$ to tensors $A_s$. Each isometry $w_s$ can be replaced with a unitary $u_s$ with a fixed input $\ket{0}$ representing a decoupled degree of freedom.
(c) Correspondence between objects in lattice and continuum TNR.
}
\label{fig:TNRcTNR}
\end{figure}

\textit{Lattice TNR.---} 
Let us briefly review the essential ingredients of the TNR algorithm on the lattice \cite{TNR,TNRyieldsMERA,TNRalgorithms, TNRlocal}. The object to be coarse-grained is a two-dimensional statistical partition function (equivalently, a discrete Euclidean path integral in two spacetime dimensions) that has been expressed as a two-dimensional tensor network, where each tensor $A$ in the network encodes local Boltzmann weights. The lattice spacing $a$ of the model serves as a short-distance cut-off. Through an intricate sequence of local manipulations of the network, which aims at removing shortly-correlated degrees of freedom, TNR effectively maps a block of four tensors $A_s$ at scale $s$ into a single tensor $A_{s+1}$ at scale $s+1$. Then space is rescaled by a factor $1/2$, so as to reset the lattice spacing $2a$ of the coarse-grained network back to the original lattice spacing $a$, see Fig. \ref{fig:TNRcTNR}(a). These general features of the method are shared with most previous tensor network coarse-graining schemes \cite{TRG,TRG2,TRG3,TRG4,TRG5,TRG6,TRG7}. What makes TNR stand out is that, thanks to the use of so-called \textit{disentanglers} $u$ and \textit{isometries} $w$ (a technology borrowed from MERA \cite{MERA1,MERA2}, see Fig. \ref{fig:TNRcTNR}(b)), it first decouples, and then eliminates, most shortly-correlated degrees of freedom from the partition function, in such a way as to generate a proper RG flow with the correct structure of fixed points.  

\begin{figure}[t]
\includegraphics[width=\columnwidth]{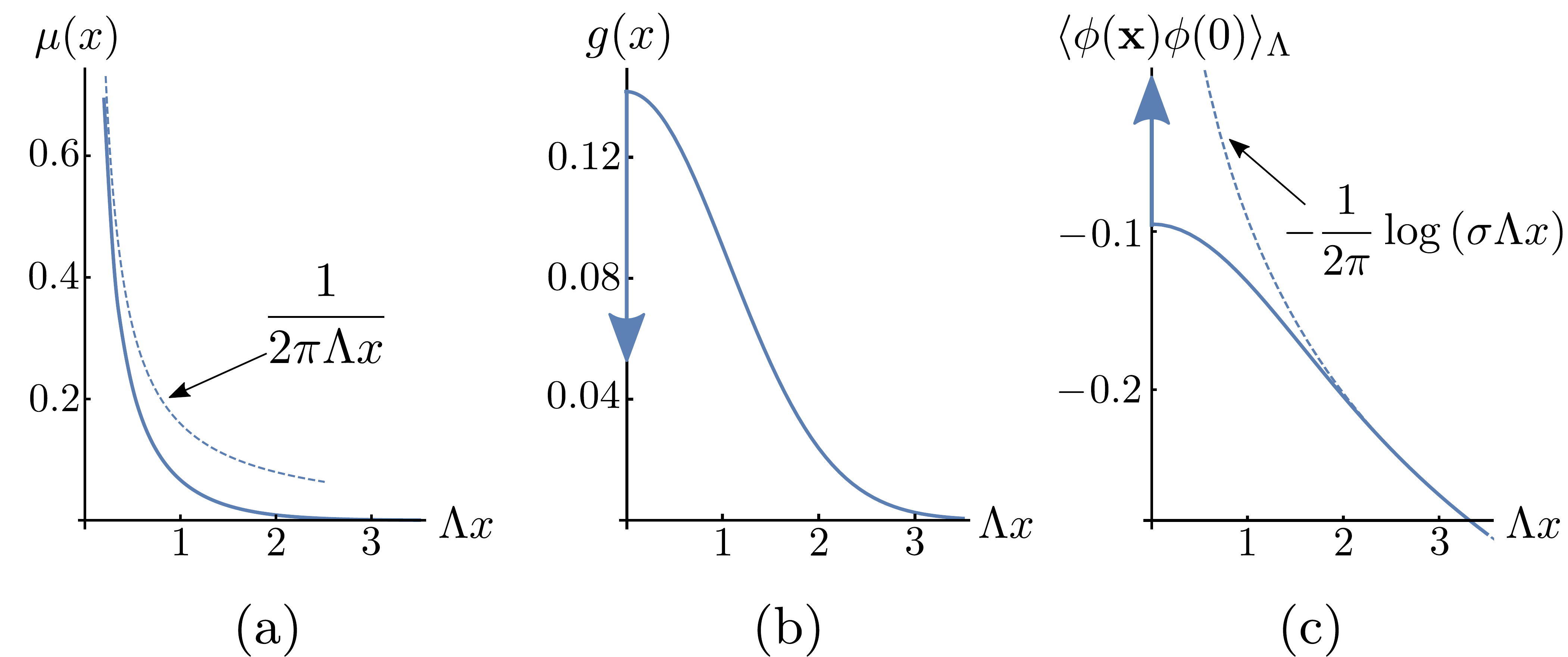}
\caption{(a) $\mu(x)$, 2D Fourier transform of $\mu(k)$ in Eq. \ref{eq:mu2D}, diverges as $1/x$ (discontinuous line) at short distances and is upper bounded by an exponentially decaying function at long distances \cite{Supplemental}. 
(b) $g(x)=-\delta(x)+\sigma\Lambda^2/(4\pi)e^{-\sigma\Lambda^2x^2/4}$, 2D Fourier transform of $g(k)$ in Eq. \ref{eq:g2D}, has a contact term at $x=0$ and decays as a Gaussian function at long distances.
(c) The correlation function $\langle \phi(\bm{x})\phi(\mathbf{0}) \rangle_{\Lambda}$, Fourier transform of $1/(k^2\mu(k)^2)$, has a delta contact term at $x=0$, and scales logarithmically (discontinuous line) at long distances \cite{Supplemental}. 
}
\label{fig:smearing_profile}
\end{figure}

\textit{Continuous partition function.---} Let us now move to a quantum field theory (QFT) in the continuum. For concreteness we consider a bosonic field $\phi(\bm{x})$ in flat $D$-dimensional Euclidean spacetime. Our object of interest is now the partition function
\begin{equation}
Z=\int [d\phi]~ e^{-S[\phi]},~~~~ S[\phi]=\int d\bm{x}~ \mathcal L \Big(\phi(\bm{x}),\Delta\phi(\bm{x}) \Big),
\end{equation} 
where the Euclidean action $S[\phi]$ is the integral of a (generally interacting) local Lagrangian density $\mathcal{L}$, which we assume to be invariant under translations and rotations. As a preliminary step, we introduce a smeared field
\begin{equation}
\phi^\Lambda(\bm{x}) \equiv \int d\bm{y}~ \mu(|\bm{x} -\bm{y}|) \phi(\bm{y}),~~~~\int d\bm{x}~ \mu(|\bm{x}|) = 1,
\label{eq:smear}
\end{equation}
where $\mu(x)\in \mathbb{R}$, with $x \equiv |\bm{x}|$, is a smearing profile invariant under O$(D)$ rotations that decays  fast to zero (\textit{e.g.} exponentially) for distances $x$ larger than a characteristic smearing length scale $1/\Lambda$, see Fig. \ref{fig:smearing_profile}(a) for an example. We then define the smeared action $S^\Lambda[\phi] \equiv S[\phi^\Lambda]$, with corresponding quasi-local Lagrangian 
\begin{equation} \label{eq:lagrangianLambda}
\mathcal{L}^\Lambda\Big(\phi(\bm{x}),\Delta\phi(\bm{x}) \Big) \equiv \mathcal{L}\Big(\phi^{\Lambda}(\bm{x}),\Delta\phi^{\Lambda}(\bm{x}) \Big),~~~~
\end{equation}
as well as the new partition function 
\begin{equation}
Z^\Lambda \equiv \int [d\phi] ~e^{-S^\Lambda[\phi]} = \int [d\phi] ~e^{-S[\phi^\Lambda]}
\end{equation}
in which fluctuations of $\phi(\bm{x})$ at distances smaller than $1/\Lambda$ have been suppressed thanks to the smearing. For instance, we expect the correlator for the sharp field $\phi(\bm{x})$, 
\begin{equation}
\langle \phi(\bm{x}) \phi(\mathbf{0})\rangle_{\Lambda} \equiv \frac{1}{Z^{\Lambda}}\int [d\phi] ~e^{-S^{\Lambda}[\phi]}~\phi(\bm{x}) \phi(\mathbf{0})
\end{equation}
to not diverge for $x \rightarrow 0$, but to tend to a constant when $\Lambda x \ll 1$, see \textit{i.e.} Fig. \ref{fig:smearing_profile}(c). The length $1/\Lambda$ plays here a role analogous to the lattice spacing $a$ in the lattice.

\textit{Continuous TNR.---} The proposed cTNR transformation proceeds through an infinitesimal change of the field
\begin{equation} \label{eq:deltaphi}
\phi(\bm{x}) \rightarrow \phi(\bm{x}) + \delta \phi(\bm{x}),~~~\delta \phi(\bm{x}) \equiv \left( L+K_s \right) \phi(\bm{x}),
\end{equation}
where $L$ is the usual \textit{rescaling} operator,
\begin{equation} \label{eq:L}
L~\phi(\bm{x} )=(-\bm{x} \cdot \nabla_{\bm{x}}-\Delta_\phi) \phi(\bm{x}),
\end{equation}
with $\Delta_\phi \equiv (D-2)/2$ the classical scaling dimension of the field $\phi(\bm{x})$, whereas the \textit{disentangling} operator $K_s$ implements a quasi-local field redefinition,
\begin{equation} \label{eq:K}
K_s \phi(\bm{x})=F\Big(s,\phi^\Lambda(\bm{x}),\Delta\phi^\Lambda(\bm{x}),\Delta^2\phi^\Lambda(\bm{x}),\dots \Big).
\end{equation}
Here $F$ is some function, not necessarily linear, of the smeared field $\phi^{\Lambda}(\bm{x})$ and its derivatives, which is invariant under translations and rotations and may also depend on the scale parameter $s$. While applying $L$ to the smeared field $\phi^{\Lambda}(\bm{x})$ shrinks its smearing length, applying the quasi-local disentangler $K_s$ is expected to restore it back to $1/\Lambda$. Transformation \ref{eq:deltaphi}-\ref{eq:K} applied to $Z^{\Lambda}$ generates an RG flow. We write symbolically \cite{Supplemental}
\begin{eqnarray}
Z_s^{\Lambda} &\equiv& \mathcal{P} e^{\int^s_0 du~ (L+K_u)}~ Z^{\Lambda} = \int [d\phi] e^{-S_s^\Lambda[\phi]}
\end{eqnarray}
where $\mathcal{P} e$ denotes a path ordered exponential. In general we should take into account the change of the integration measure when computing the evolution of the action $S_s^\Lambda[\phi]$. Since both $L$ and $K_s$ act (quasi-)locally, it should be possible to write the new action as an integral of a quasi-local Lagrangian density:
\begin{equation}
S_s^\Lambda[\phi]=\int d\bm{x}~ \mathcal L_s^\Lambda \Big(\phi(\bm{x}),\Delta\phi(\bm{x}),\Delta^2\phi(\bm{x}),\dots \Big).
\end{equation}

Off criticality, we expect a flow of $\mathcal L_s^\Lambda$ with $s$ towards some massive fixed point Lagrangian. At a critical point, instead, we expect a flow towards an unstable fixed point Lagrangian $\mathcal L_{\star}^\Lambda$ corresponding to a (smeared version of) a CFT. This will be characterized by a spectrum of quasi-local scaling operators $\mathcal{O}^{\Lambda}_{\alpha}(\bm{x})$. For instance, in $D=2$ dimensions the latter are solutions to
\begin{eqnarray} \label{eq:LKstar}
(L+K_{\star})~O^\Lambda_\alpha(\mathbf{0})&=&-\Delta_\alpha ~O^\Lambda_\alpha(\mathbf{0}),\\
R~O^\Lambda_\alpha(\mathbf{0})&=&s_\alpha ~O^\Lambda_\alpha(\mathbf{0}), \label{eq:R}
\end{eqnarray}
where $\Delta_\alpha$ and $s_\alpha$ are the scaling dimension and conformal spin of $O^\Lambda_\alpha(x)$, $K_{\star}$ is the fixed-point disentangling operator, and $R$ is the generator of rotations in the Euclidean plane, $R~\phi(\bm{x})= (x_1 \partial_{x_2} - x_2\partial_{x_1})~ \phi(\bm{x})$.

\textit{Continuum versus lattice.---} As illustrated in Fig. \ref{fig:TNRcTNR}(c), we can think of $e^{-d\bm{x} \mathcal{L}_s^{\Lambda}}$ as the continuum counterpart of a tensor $A_s$ in the network representing a partition function on the lattice. Then $L + K_s$ implement in the continuum the equivalent of a TNR coarse-graining transformation on the lattice, with the disentangler $K_s$ being the continuum version of the disentanglers $u_s$ and isometries $w_s$ on the lattice. Indeed, both the continuum $K_s$ and the lattice $u_s$ and $w_s$ implement a (quasi-)local reorganization of the degrees of freedom that aims to decouple from the partition function those that are shortly correlated, that is, correlated at lengths on the order $a\sim 1/\Lambda$. However, while on the lattice each step of TNR implements a \textit{discrete} change of scale $s \rightarrow s+1$ and the disentanglers $u_s$ and isometries $w_s$ are used to \textit{completely} decouple half of the lattice degrees of freedom, in the continuum TNR implements instead a \textit{continuous} change of scale $s$, during which the disentangler $K_s$ \textit{gradually} decouples field degrees of freedom. 

\textit{Example: free boson in two dimensions.---} Some features of the above general proposal for interacting field theories can be well illustrated using the simplified scenario of free fields. For concreteness, here we consider a free boson in $D=2$ spacetime dimensions. The Euclidean action is given by
\begin{eqnarray}\label{eq:S2D}
S[\phi]&=&\frac{1}{2} \int d \bm{x} \left( -\phi(\bm{x})\Delta\phi(\bm{x}) +m^2 \phi(\bm{x})^2 \right)\\
&=& \frac{1}{2} \int \frac{d\bm{k}}{(2\pi)^2}~ (k^2+m^2) \phi(\bm{k})\phi(-\bm{k}),
\end{eqnarray} 
where $\phi(\bm{k}) \equiv \int d\bm{x}~\phi(\bm{x})e^{-i\bm{k} \cdot \bm{x}}$ is a Fourier mode. Although cTNR is a real-space renormalization scheme, for free fields it is insightful to work in momentum space. The following derivations require performing standard Gaussian integrations and Fourier transforms, as detailed in \cite{Supplemental}. The momentum-space two-point correlator reads 
\begin{equation} \label{eq:corrMomentum}
\langle \phi(\bm{k}) \phi(-\bm{k}) \rangle 
= \frac{1}{k^2+m^2},
\end{equation} 
leading to a real-space correlator $\langle \phi(\bm{x}) \phi(\mathbf{0}) \rangle$ that diverges at short distances, $x\rightarrow 0$. For instance, for $m=0$,
\begin{equation} \label{eq:corrReal}
\langle \phi(\bm{x}) \phi(\mathbf{0}) \rangle = -\frac{1}{2\pi} \log (x) + \mbox{const.}~~~~~~\mbox(m=0).
\end{equation} 

Instead the smeared action $S^\Lambda[\phi] \equiv S[\phi^\Lambda]$ reads 
\begin{eqnarray} \label{eq:S2DLam}
S^\Lambda[\phi] &=& \frac{1}{2} \int d \bm{x} \Big( -\phi^\Lambda(\bm{x})\Delta\phi^\Lambda(\bm{x}) +m^2 \phi^\Lambda(\bm{x})^2 \Big)\\
&=&\frac{1}{2} \int \frac{d \bm{k}}{(2\pi)^2}~ (k^2+m^2)\mu(k)^2 \phi(\bm{k})\phi(-\bm{k}),
\end{eqnarray}
and leads to the sharp field correlator 
\begin{eqnarray}
\langle \phi(\bm{k}) \phi(-\bm{k}) \rangle_{\Lambda} 
= \frac{1}{(k^2+m^2)}\frac{1}{\mu(k)^2}. \label{eq:corrMomentumLam}
\end{eqnarray}
Above we have used the Fourier transform of Eq. \ref{eq:smear}, $\phi^\Lambda(\bm{k})=\mu(k)\phi(\bm{k})$. Since the smearing profile $\mu(x)$ is real and rotation invariant, so is $\mu(k)$. We further constrain $\mu(k)$ with two requirements. First, we would like $\langle \phi(\bm{k}) \phi(-\bm{k}) \rangle_{\Lambda}$ to coincide with $\langle \phi(\bm{k}) \phi(-\bm{k}) \rangle$ for $k \ll \Lambda$, so that the smeared field theory reproduces the large distance physics of the original field theory. Second, we would like to remove the short-distance divergence in $\langle \phi(\bm{x}) \phi(\mathbf{0}) \rangle$ (see e.g. Eq. \ref{eq:corrReal}), which demands that $\langle \phi(\bm{k}) \phi(-\bm{k}) \rangle_{\Lambda}$ tend to a constant $1/\Lambda^2$ sufficiently fast for $k\gg \Lambda$. Accordingly we will require \cite{Supplemental}: 
\begin{eqnarray}
&&~~~\mu(k) = \left\lbrace \begin{array}{cl}
1  & \mbox{for}~ k/\Lambda\rightarrow 0, \\
\Lambda/k  & \mbox{for}~ k/\Lambda\rightarrow \infty,
\end{array}\right. \nonumber\\
&&\left|\int_\Lambda^{\infty} \left(\frac{1}{(k^2+m^2)}\frac{1}{\mu(k)^2} - \frac{1}{\Lambda^2} \right)\right| < \infty.\label{eq:constraints}
\end{eqnarray} 

\textit{Free boson cTNR.---} For a free theory, we can use a disentangling operator $K_s$ linear in $\phi(\bm{x})$, 
\begin{equation}
K_s ~\phi(\bm{x})=\int  d\bm y~g (s,|\bm{x} -\bm{y}|) ~\phi(\bm{y}),
\end{equation}
or $K_s \phi(\bm{k})=g(s,k) \phi(\bm{k})$ in momentum space. Notice that $K_s$ is built to be invariant under both translations and rotations, since for any $s$, $g(s,|\bm{x} -\bm{y}|)$ is only a function of $|\bm{x} -\bm{y}|$. In analogy with lattice TNR, where disentanglers $u_s$ and isometries $w_s$ act locally on a region of linear size $a$, we further require $g(s,x)$ to be a quasi-local function of $x$ with characteristic length scale $1/\Lambda$, see \textit{i.e.} Fig. \ref{fig:smearing_profile}(b). For $s=0$, $L+K_s$ acts on $S^{\Lambda}[\phi]$ as \cite{Supplemental}
\begin{eqnarray}
&&(L+K_s) ~S^\Lambda[\phi]\\
&=&\int \frac{d\bm{k}}{(2\pi)^2}  \Big\{  k^2 \mu(k)\Big [\big (-k\partial_k+g(k)\big )\mu(k) \Big ]\\
&+&m^2 \mu(k) \left[\left(-k\pa_k+1+g(k)\right)\mu(k) \Big ] \right\}\phi(\bm{k})\phi(-\bm{k}),~~
\end{eqnarray}
with $g(k)\equiv g(0,k)$. It follows that, in the massless case $m=0$, 
the action $S^{\Lambda}[\phi]$ is invariant if and only if
\begin{equation}
g(k)=\frac{k\pa_k \mu(k)}{\mu(k)}. \label{eq:diff}
\end{equation}
Let $K_{\star}$ denote the fixed-point entangler (that is, with $g(k)$ obeying \ref{eq:diff} ) and let $S^{\Lambda}_{\star}[\phi]$ denote the massless action, \textit{i.e.} $(L+K_{\star}) ~S^{\Lambda}_{\star}[\phi]=0$. It also follows \cite{Supplemental} that 
\begin{equation} \label{eq:LKphi}
(L+K_{\star})~\phi^\Lambda(\mathbf{0})= 0,
\end{equation}
which implies that the effect of the rescaling operator $L$ on the smeared field $\phi^{\Lambda}(\mathbf{0})$ (namely the shrinking of its smearing profile $\mu(x)$) is exactly compensated by that of the fixed-point disentangler $K_{\star}$ (which re-expands the smearing through a quasi-local field redefinition). Finally, as a concrete example, the pair of functions
\begin{eqnarray}
\mu(k)&=&\f{\Lambda}{k}\textrm{exp}\left( \f 1 2 \textrm{Expi}\left(-\f{k^2}{\sigma \Lambda^2}\right) \right), \label{eq:mu2D}\\
g(k)&=&-1+\exp \left(-\frac{k^2}{\sigma \Lambda^2}\right),
\label{eq:g2D}
\end{eqnarray}
where $\textrm{Expi}(x)$ is the exponential integral function and $\sigma = e^{\gamma} \approx 1.78$ (with $\gamma$ Euler's constant), fulfill the constraints \ref{eq:constraints} and \ref{eq:diff} while their Fourier transforms $\mu(x)$ and $g(x)$, depicted in Fig. \ref{fig:smearing_profile}(a,b), are quasi-local with characteristic length $1/\Lambda$ \cite{Supplemental}.  Fig. \ref{fig:smearing_profile}(c) shows the resulting correlator $\langle \phi(\bm{x}) \phi(\mathbf{0})\rangle_{\Lambda}$, which is UV-finite.

\textit{RG flow and critical fixed point.---}  
Applying now the above fixed-point disentangler $K_{\star}$ to the action $S^{\Lambda}[\phi]$ for $m\not = 0$ \cite{Kstar} results in a scale-dependent action
\begin{equation}
S_s^\Lambda[\phi] = \frac{1}{2} \int d \bm{x} \Big( -\phi^\Lambda(\bm{x})\Delta\phi^\Lambda(\bm{x}) + m(s)^2 \phi^\Lambda(\bm{x})^2 \Big)
\end{equation}
where the mass $m(s) \equiv m e^{s}$ grows exponentially with the RG scale $s$. Thus we have recovered the well-known RG flow of a massive free boson towards its infinite mass fixed point. 

Returning to the critical point, with fixed-point Lagrangian $\mathcal{L}^{\Lambda}_{\star} \equiv -\frac{1}{2}\phi^{\Lambda}(\bm{x})\Delta\phi^{\Lambda}(\bm{x})$, it can be shown that the quasi-local scaling operators $\mathcal{O}^{\Lambda}_{\alpha}(\bm{x})$, \textit{cf.} Eqs. \ref{eq:LKstar}-\ref{eq:R}, are in one to one correspondence with the local scaling operators $\mathcal{O}_{\alpha}(\bm{x})$ of the free boson CFT and can obtained by smearing them \cite{Supplemental}. This observation is analogous to that in Ref. \cite{Qi}. For example, the right moving field $\partial \phi(\bm{x})\equiv (\partial_{x_1}-i\partial_{x_2})\phi(\bm{x})$ is a CFT scaling operator with scaling dimension $\Delta_{\pa\phi}=1$ and conformal spin $s_{\pa\phi}=1$, satisfying $L~\partial\phi(\mathbf{0})=-\partial\phi(\mathbf{0})$ and $R~\partial\phi(\mathbf{0})=\partial \phi(\mathbf{0})$. By smearing those expressions we readily find the corresponding scaling operator $\partial\phi^\Lambda(\bm{x})$:
\begin{eqnarray}
(L+K_\star)~\partial \phi^\Lambda(\mathbf{0})&=& -\partial \phi^\Lambda(\mathbf{0}), \\
R~\partial\phi^\Lambda(\mathbf{0}) &=& \partial \phi^\Lambda(\mathbf{0}),
\end{eqnarray}
with the exact same scaling dimension and conformal spin. We can similarly recover the operator product expansion and central charge of the original CFT \cite{CFT1,CFT2,CFT3}, and therefore extract all of its conformal data \cite{Supplemental}.
 
\textit{Discussion.---} In this \letter we have proposed an extension of the TNR formalism \cite{TNR,TNRyieldsMERA,TNRalgorithms, TNRlocal} to quantum fields in the continuum and demonstrated with a free boson that, as on the lattice, continuous TNR generates a proper RG flow, including a critical fixed point from which one can extract the universal critical properties (conformal data) of the phase transition. The exact preservation of translation and rotation symmetry, accomplished through the use of explicitly symmetric smearing function $\mu$ and disentangling operator $K_s$, demonstrates the possibility of preserving such symmetries in a real space RG approach. It is also expected to lead to increased numerical robustness and reduced computational costs with respect to lattice TNR. 

Importantly, an actual cTNR algorithm for interacting QFTs is currently still missing. However, based on the success of TNR and related algorithms for interacting models on the lattice \cite{TNR,TNRyieldsMERA,TNRalgorithms, TNRlocal,loopTNR, TNRplus, skeleton, giltTNR}, it is reasonable to expect that one such algorithm will be eventually developed. We envisage that cTNR will then represent a powerful alternative to Wilsonian RG methods \cite{Wilson}. Recall that the latter operate in \textit{momentum space} and are based on sequentially \textit{integrating out} thin shells of modes with large momentum. We emphasize that cTNR, a \textit{real space} method, operates in a fundamentally different way by \textit{decoupling out} shortly-correlated degrees of freedom through the use of a quasi-local field redefinition.

Our proposal parallels the development of the cMERA, put forward by Haegeman, Osborne, Verschelde, and Verstraete in Ref. \cite{cMERA}. As cMERA \cite{cMERA,Qi,Adrian}, the cTNR formalism is based on smeared fields and is only well understood for free particle QFTs. Moreover, at criticality both cMERA \cite{Qi} and cTNR \cite{Supplemental} can be seen to realize conformal symmetry quasi-locally. However, even though TNR and MERA are tightly related on the lattice \cite{TNRyieldsMERA},
in the continuum there exist a clear divide between the two formalism. Indeed, in cMERA the fields are only smeared in the space direction, whereas in cTNR the smearing is isotropic in Euclidean spacetime. As a result, in cTNR it is unclear how to even define the Hilbert space attached to a constant time slice in which cMERA would represent a many-body wavefunctional \cite{cTNRanisotropic}. 
Finally, while awaiting the development of a cTNR algorithm for interacting QFTs, we hope that cTNR will become a useful framework for holographic studies, thus following the path of both lattice TNR  \cite{TNRholo1, TNRholo2, TNRholo3, TNRholo4} and cMERA \cite{cMERAholo1, cMERAholo2, cMERAholo3, cMERAholo4, cMERAholo5, cMERAholo6, cMERAholo7, cMERAholo8, cMERAholo9}.

A. F.-R. acknowledges support from the La Caixa Graduate Fellowship Program. The authors acknowledge support from the Simons Foundation (Many Electron Collaboration) and from the Natural Sciences and Engineering Research Council of Canada (NSERC) through a Discovery Grant. This research was supported in part by Perimeter Institute for Theoretical Physics. Research at Perimeter Institute is supported by the Government of Canada through the Department of Innovation, Science and Economic Development Canada and by the Province of Ontario through the Ministry of Research, Innovation and Science.

\textit{Note:} After completion of the results presented in this manuscript, we learned of the recent paper \textit{Continuous tensor network states for quantum fields} \cite{Tilloy}, which also discusses the extension to the continuum of lattice tensor networks. The focus there is on (tensor network representations of) wavefunctions, while the focus in our paper is instead on partition functions/Euclidean path integrals. It would certainly be of interest to explore possible connections between the two proposals.


\section{APPENDICES}

\section{quasilocal smearing function}
In this section we examine the constraints of the smearing function $\mu(\bm x, \bm y)$ and illustrate a concrete example which fulfills the constraints. We will have four constraints. (i) The smearing function $\mu (\bm x ,\bm y)$ is translation and rotation invariant, which means that it only depends on $|\bm x -\bm y|$. (ii) To (quasi)preserve the local structure of the continuous partition function, we require the smearing function $\mu( x)$ to be quasilocal, in the sense that it is upper bounded by an exponentially decaying function at long distances. (iii) The smeared fields should be normalized such that they have the same behaviors as in the original field theory at long distances. This demands that $\int d\bm x~ \mu( |\bm x|)=1$, or equivalently $\mu( k= 0)=1$. (iv) We demand that there are only finite fluctuations at short distances, that is,
\beq
\left\lvert \lim_{\bm x\to 0} \la \phi(\bm x) \phi(\bm 0) \ra _\Lambda \right\rvert <\infty,
\eeq
where $\la \phi(\bm x) \phi(\bm 0) \ra _\Lambda$ represents the correlation function for the smeared action $S^\Lambda[\phi]$ determined by the smearing function $\mu(x)$. Assuming $\la \phi(\bm k)\phi(-\bm k) \ra_\Lambda$ tends to a constant $\f{1}{\Lambda^2}$, which results in a contact term in the correlation function  $\la \phi(\bm x)\phi(\bm 0) \ra_\Lambda$ in real space, constraint (iv) becomes
\beq
\left\lvert\int d\bm k~ \Big( \f{1}{k^2+m^2}\f{1}{\mu(k)^2}-\f{1}{\Lambda^2} \Big)\right\rvert<\infty.
\eeq
Here the integral should be understood as the Hadamard finite-part integral if divergence occurs in the neighborhood of $k=0$. In that case, the condition can also be written as 
\beq
\left\lvert\int_{k>\Lambda} d\bm k~ \Big( \f{1}{k^2+m^2}\f{1}{\mu(k)^2}-\f{1}{\Lambda^2} \Big)\right\rvert<\infty.
\label{constraint3}
\eeq

An example of $\mu(x)$ that fulfills the four constraints is given by the 2D Fourier transform of the following:
\beq
\mu(k)=\f{\Lambda}{k}\textrm{Exp}\left( \f 1 2 \textrm{Expi}\left(-\f{k^2}{\sigma \Lambda^2}\right)\right).
\label{defmu}
\eeq
Here $\sigma=e^\gamma\approx 1.78$, where $\gamma$ is Euler's constant. 
Next we show that constraints (i), (ii) and (iii) are satisfied. In the following section we will show that constraint (iv) is also fulfilled. 

Constraint (i) is fulfilled because $\mu(k)$ only depends on $|\bm k|$. Constraint (iii) can be verified by plugging in the expansion 
\beq
\textrm{Expi}(x)=\gamma+\log x+O(x^2),  ~~~~~~   \textrm{       as }x\to 0,
\eeq
in Eq. \ref{defmu}, which yields
\beq
\mu(k)=1-\f{k^2}{2\sigma\Lambda^2}+O(k^4),  ~~~~~~   \textrm{       as }k\to 0,
\eeq
and in particular $\mu(k=0)=1$.

%
\begin{figure}[h]
\includegraphics[width=3.5in]{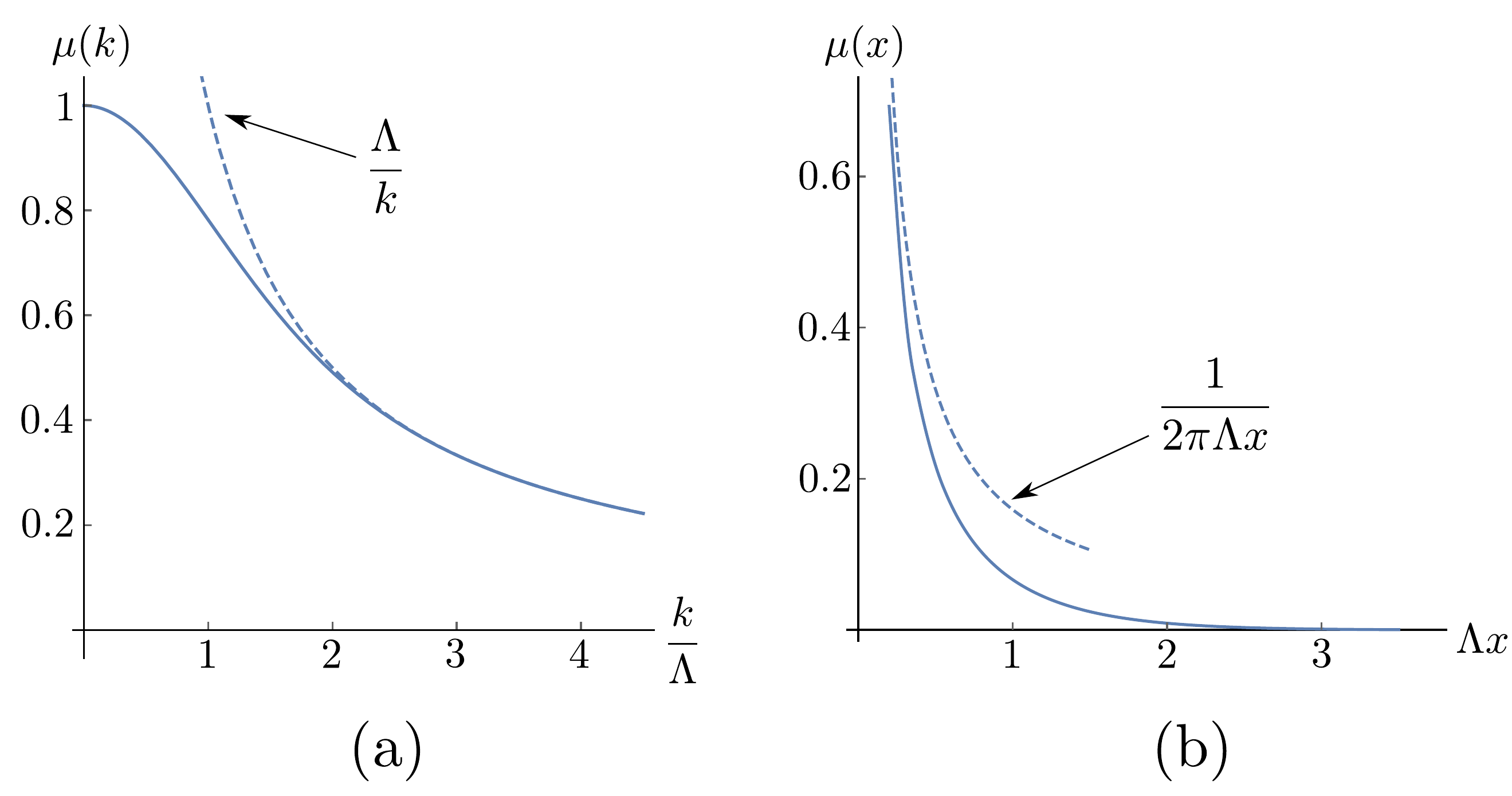}
\caption{(a) The smearing function $\mu(k)$ in momentum space. The dashed line is $\f{\Lambda}{k}$, which represents the asymptotic behavior of $\mu(k)$ for large $k$. (b) The smearing function $\mu(x)$ in real space. The dashed line is $\f{1}{2\pi\Lambda x}$, which represents the asymptotic behavior of $\mu(x)$ for small $x$.}
\label{mu}
\end{figure}

Constraint (ii) will follow from the analytical study of the asymptotic behavior of $\mu(x)$ for large $x$, which can be inferred from its 2D Fourier transform $\mu(k)$. We use the same strategy as in the appendix of \cite{Qi}. Notice that $\mu(k)$ satisfies the differential equation
\beq
k\pa_k \mu(k)=g(k)\mu(k),
\eeq
where $g(k)=-1+e^{-\f{k^2}{\sigma \Lambda^2}}$. Applying a 2D Fourier transform, we get the equation in real space,
\beq
(-\bm x \cdot \nabla -1)\mu(\bm x)=\f{\sigma \Lambda^2}{4\pi}\int d\bm y ~ e^{-\f 1 4 \sigma \Lambda^2 y^2}\mu(\bm x -\bm y).
\eeq
We define a new variable $u(\bm x)$:
\beq
\mu(\bm x)=e^{-u(\bm x)}.
\eeq
Then the differential equation becomes
\beq
\bm x \cdot \nabla u(\bm x) -1=\f{\sigma \Lambda^2}{4\pi}\int d\bm y ~ e^{-\f 1 4 \sigma \Lambda^2 y^2-u(\bm x -\bm y)+u(\bm x)}.
\eeq
We then expand $u(\bm x -\bm y)$ around $\bm x$,
\beq
u(\bm x -\bm y)\approx u(\bm x )-\bm y \cdot \nabla u(\bm x).
\eeq
In consequence, the right hand side of the differential equation is approximated by a Gaussian integral,
\beqa
\blgn
&\f{\sigma \Lambda^2}{4\pi}\int d\bm y ~ e^{-\f 1 4 \sigma \Lambda^2 y^2-u(\bm x -\bm y)+u(\bm x)} \\
&\approx \f{\sigma \Lambda^2}{4\pi}\int d\bm y ~ e^{-\f 1 4 \sigma \Lambda^2 y^2+\bm y \cdot \nabla u(\bm x)}\\
&=e^{(\nabla u(\bm x))^2/(\sigma \Lambda^2)}.
\elgn
\eeqa
Since $\mu(\bm x)$ is rotation invariant, it is only a function of $x$, and hence so is $u(\bm x)$. Then we have
\beq
x u'(x)-1=e^{u'(x)^2/(\sigma\Lambda^2)}.
\eeq
The approximate solution can be found in the large $x$ limit:
\beq
u'(x)=\Lambda \sqrt{\sigma~ \textrm{log}\Lambda x}~(1+o(1)),  \textrm{     as }x\to \infty.
\eeq 
Performing the integration we get
\beq
u(x)=\Lambda x \sqrt{\sigma~ \textrm{log}\Lambda x}~(1+o(1)),  \textrm{     as }x\to \infty.
\eeq
Finally we obtain the asymptotic behavior of $\mu(x)$ for large $x$,
\beq
\mu(x)=e^{-\Lambda x\sqrt{\sigma~ \textrm{log}\Lambda x}~(1+o(1))},  \textrm{     as }x\to \infty.
\eeq 
It is then clear that $\mu(x)$ can be upper bounded by an exponentially decaying function in the limit of large $x$. Therefore our proposed smearing function fulfills constraint (ii).

Additionally, the asymptotic behavior of $\mu(x)$ for small $x$ can also be obtained from its Fourier transform. To do so, we divide $\mu(k)$ into two pieces $\mu^{(1)}(k)+\mu^{(2)}(k)$:
\beq
\mu^{(1)}(k)\equiv\f \Lambda k,
\eeq
\beq
\mu^{(2)}(k)\equiv\f{\Lambda}{k}\textrm{Exp}\Big( \f 1 2 \textrm{Expi}(-\f{k^2}{\sigma \Lambda^2})                \Big)-\f \Lambda k.
\eeq
The 2D Fourier transform of $\mu^{(1)}(k)$ is $\f{1}{2\pi \Lambda x}$, which is computed analytically, and the 2D Fourier transform of $\mu^{(2)}(k)$ can be shown to be finite around $x=0$. Therefore, the asymptotic behavior of $\mu(x)$ at $x=0$ is
\beq
\mu(x)= \f{1}{2\pi \Lambda x}+O(1),  \textrm{     as }x\to 0.
\eeq


\section{correlation function}
In this section we study the behavior of the correlation function $\la \phi(\bm x)\phi(\bm 0) \ra_\Lambda$ for the smeared free boson action $S^\Lambda[\phi]$. For concreteness, the analysis of the short and long distance behavior of the correlator will focus on the massless case. At short distances, it can be shown to be finite, thus fulfilling constraint (iv) from the previous section. At long distances, it approaches the correlation function $\la \phi(\bm x)\phi(\bm 0) \ra$ from the CFT.

For the 2D free boson theory in Euclidean spacetime, the action is given by
\beqa
\blgn
S[\phi]&=\f 1 2 \int d \bm x \left( -\phi(\bm x)\Delta\phi(\bm x) +m^2 \phi(\bm x)^2 \right)\\
&=\f 1 2 \int \f{d \bm k}{(2\pi)^2}~ (k^2+m^2) \phi(\bm k)\phi(-\bm k).
\elgn
\eeqa
To compute the correlation function $\la \phi(\bm k)\phi(-\bm k) \ra$, we introduce a source term in the action:
\beqa
\blgn
S[\phi,\lambda]&=\f 1 2 \int d \bm x ~\Big( -\phi(\bm x)\Delta\phi(\bm x) +m^2 \phi(\bm x)^2 \Big)\\
&~~~+\int d \bm x~ \lambda(x) \phi(\bm x)\\
&=\f 1 2 \int \f{d \bm k}{(2\pi)^2}~ (k^2+m^2) \phi(\bm k)\phi(-\bm k)\\
&~~~+ \int \f{d \bm k}{(2\pi)^2}~  \lambda(\bm k)\phi(-\bm k).
\elgn
\eeqa
The partition function is given by the path integral:
\beqa
\blgn
Z(\lambda)&\equiv\int [d\phi]e^{-S[\phi,\lambda]}\\
&=Z(0)e^{\f 1 2 \int \f{d \bm k}{(2\pi)^2} (k^2+m^2)^{-1}\lambda(\bm k)\lambda(-\bm k)}.
\elgn
\eeqa
The correlation function can then be computed as follows:
\beqa
\blgn
&\la \phi(\bm k)\phi(-\bm k) \ra\\
&=\f{\int [d\phi]e^{-S[\phi]}\phi(\bm k)\phi(-\bm k)}{\int [d\phi]e^{-S[\phi]}}\\
&=\left. Z(0)^{-1}\f{\pa }{\pa \lambda(\bm k)}\f{\pa }{\pa \lambda(-\bm k)}\int [d\phi]e^{-S[\phi,\lambda]}\right\lvert_{\lambda(\bm k)=0}\\
&=\left. Z(0)^{-1}\f{\pa }{\pa \lambda(\bm k)}\f{\pa }{\pa \lambda(-\bm k)}Z(\lambda)\right\lvert_{\lambda(\bm k)=0}\\
&=\f{1}{k^2+m^2}.
\elgn
\eeqa
In the massless case $m=0$, $\la \phi(\bm k)\phi(\bm -k) \ra=\f{1}{k^2}$. Its 2D Fourier transform produces the real-space correlator,
\beqa
\blgn
\la \phi(\bm x)\phi(\bm 0) \ra&=\int \f{d \bm k}{(2\pi)^2} \la \phi(\bm k)\phi(-\bm k) \ra e^{i\bm k \cdot \bm x}\\
&=-\f{1}{2\pi}\log(x)+\textrm{const.}
\elgn
\label{log_const}
\eeqa

\begin{figure}[h]
\includegraphics[width=2.8in]{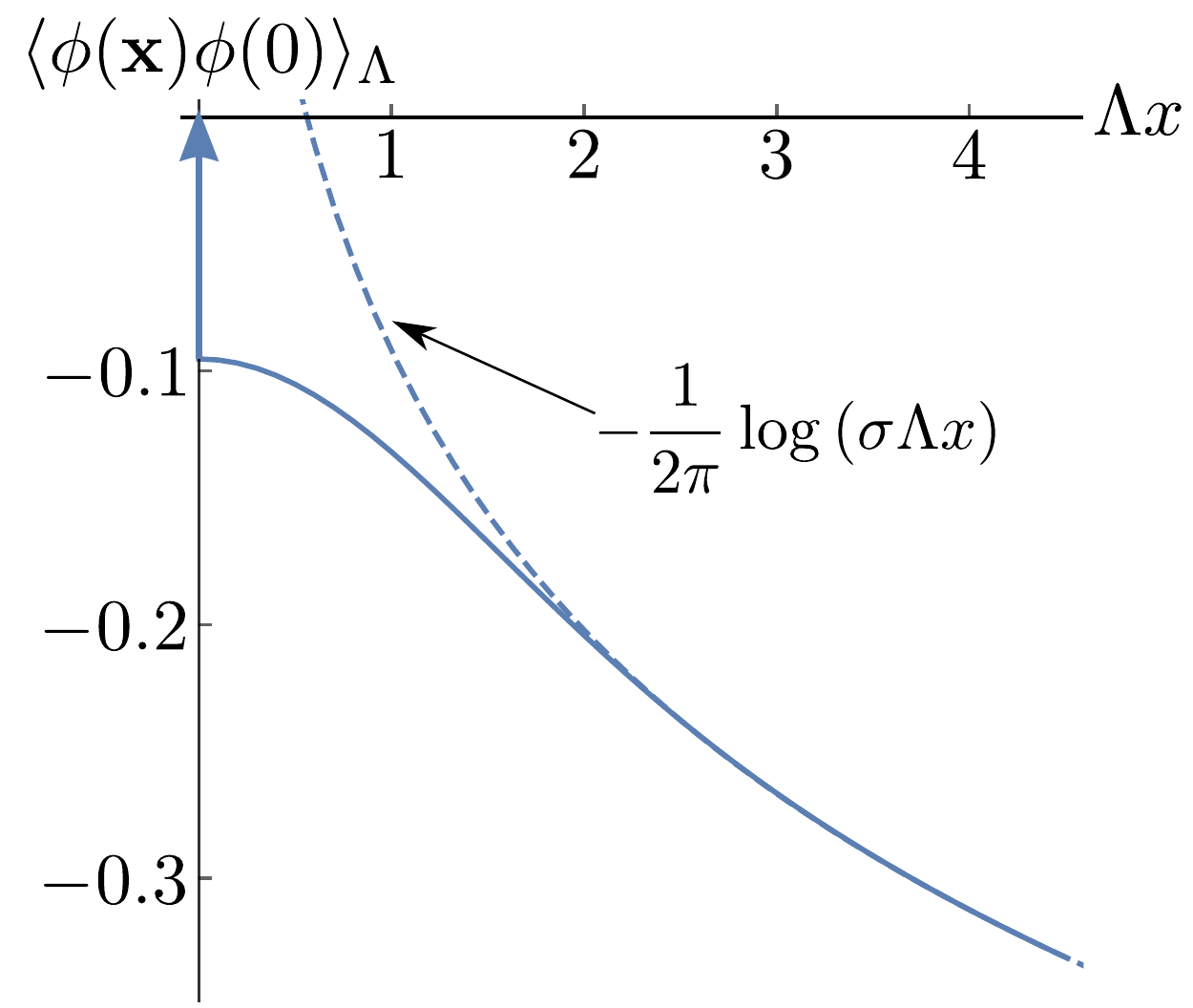}
\caption{Correlation function $\la \phi(\bm x)\phi(\bm 0)    \ra_\Lambda$. The arrow represents a delta function $\f{1}{\Lambda^2}\delta(\bm x)$. The dashed line is $-\f{1}{2\pi} \log(\sigma \Lambda x)$, which represents the asymptotic behavior of the correlation function for large $x$, and matches the CFT correlation formula Eq. \ref{log_const}}
\label{phiphix}
\end{figure}

Similarly we obtain the correlation function for the smeared action:
\beqa
\blgn
\la \phi(\bm k)\phi(-\bm k) \ra_\Lambda&=\f{\int [d\phi]e^{-S[\phi^\Lambda]}\phi(\bm k)\phi(-\bm k)}{\int [d\phi]e^{-S[\phi^\Lambda]}}\\
&=\f{1}{(k^2+m^2)\mu(k)^2}.
\elgn
\eeqa
In the massless case, $\la \phi(\bm k) \phi(-\bm k) \ra_\Lambda=\f{1}{k^2 \mu(k)^2}$. Note that $\f{1}{k^2 \mu(k)^2}\sim \f{1}{k^2}$ as $k\to 0$, and $\f{1}{k^2 \mu(k)^2}\sim \f{1}{\Lambda^2}$ as $k\to \infty$. To compute the numerical value of the Fourier transform, we subtract the divergent part $\f{1}{k^2}+\f{1}{\Lambda^2}$, which has analytical Fourier transform $-\f{1}{2\pi} \log(\sigma \Lambda x)+\f{1}{\Lambda^2}\delta(\bm x)$ (where we have made a concrete choice of the arbitrary additive constant in Eq. \ref{log_const}). We then perform the numerical Fourier transform to the rest and add it to the analytical part. The final result is shown in Fig. \ref{phiphix}. As we can see, at short distances the correlation function $\la \phi(\bm x)\phi(\bm 0) \ra_\Lambda$ is finite apart from a contact term, and at long distances, $\la \phi(\bm x)\phi(\bm 0) \ra_\Lambda$ asymptotically approaches the correlation function in the boson CFT. One can easily verify Eq. \ref{constraint3}, which also proves that the correlation function is finite at short distances. For example, in the massless case $m=0$, $\left\lvert\int_{k>\Lambda} d\bm k~ \Big(\f{1}{k^2\mu(k)^2}-\f{1}{\Lambda^2} \Big)\right\rvert \approx 1.855.$

\section{RG flow generated by $L+K_s$}

In this section, we investigate how the action $S^\Lambda[\phi]$ and the smeared field $\phi^\Lambda(\bm x)$ evolve along the RG flow generated by $L+K_s$. Recall the definition of $L$:
\beq
L\phi(\bm x )=(-\bm x \cdot \nabla_{\bm x}-\Delta_\phi) \phi(\bm x).
\eeq
For the 2D free boson theory, $\phi(\bm x)$ has classical scaling dimension $\Delta_\phi=0$. Therefore, the above equation becomes
\beq
L\phi(\bm x )=-\bm x \cdot \nabla_{\bm x} \phi(\bm x).
\eeq
In momentum space it reads
\beq
L\phi(\bm k )=(\bm k \cdot \nabla_{\bm k} +2)\phi(\bm k).
\eeq
The action of $K_s$ in real space is given by
\beq
K_s \phi(\bm x)=\int d\bm y~ g(s,|\bm x -\bm y|) \phi(\bm y).
\label{K_k}
\eeq
In momentum space, it reads
\beq
K_s \phi(\bm k)=g(s,k) \phi(\bm k).
\label{K_k}
\eeq
In flat spacetime, the path integration measure $[d\phi]$ can be decomposed as a product of measures for individual momentum modes, that is, $[d\phi]=\prod_{\bm k} d\phi(\bm k)$. Eq. \ref{K_k} implies that $K$ acts diagonally in momentum space, and therefore only changes the integration measure by a constant factor. We further assume that $L$ also leaves the integration measure invariant up to a constant factor. Omitting both of these constant factors, we can generate an RG flow by directly applying $L+K_s$ to the action $S^\Lambda[\phi]$. In order to alleviate notation, we only consider $K_s$ for $s=0$, and define $K\equiv K_0$, $g(k)\equiv g(0,k)$.
\beqa
\blgn
&(L+K)S^\Lambda[\phi]\\
&=\f 1 2 \int \f{d \bm k}{(2\pi)^2} (k^2+m^2)\mu(k)^2 \Big[ (L+K)\phi(\bm k)\phi(-\bm k)\\
&~~~+\phi(\bm k)(L+K)\phi(-\bm k) \Big ]\\
&=\f 1 2 \int \f{d \bm k}{(2\pi)^2}  \Big [ (k^2+m^2)\mu(k)^2 \\
&~~~\Big(\bm k \cdot \nabla_{\bm k} +4+2g(k)\Big)\Big(\phi(\bm k)\phi(-\bm k)\Big) \Big]\\
&=\f 1 2 \int \f{d \bm k}{(2\pi)^2} \Big[ \Big(-\bm k \cdot \nabla_{\bm k} +2+2g(k)\Big)\\
&~~~\Big((k^2+m^2)\mu(k)^2 \Big)\phi(\bm k)\phi(-\bm k)\Big]\\
&=\f 1 2 \int \f{d \bm k}{(2\pi)^2} \Big[ k^2\Big(-\bm k \cdot \nabla_{\bm k}+2g(k)\Big)\Big(\mu(k)^2 \Big)\\
& ~~~+ m^2\Big(-\bm k \cdot \nabla_{\bm k}+2+2g(k)\Big)\Big(\mu(k)^2 \Big)\Big]\phi(\bm k)\phi(-\bm k)\\
&=\int \f{d\bm k}{(2\pi)^2}  \Big\{  k^2 \mu(k)\Big [\big (-k\pa_k+g(k)\big )\mu(k) \Big ]\\
&~~~+m^2 \mu(k) \Big [\big (-k\pa_k+1+g(k)\big )\mu(k) \Big ] \Big\}\phi(\bm k)\phi(-\bm k).
\elgn
\eeqa
Similarly, we can obtain the change of the smeared field $\phi^\Lambda(\bm x)$ under the action of $L+K$. For simplicity, we only consider $\phi^\Lambda(\bm x)$ for $\bm x=\bm 0$.
\beqa
\blgn
&(L+K_s)\phi^\Lambda(\bm 0)\\
&=\int \f{d \bm k}{(2\pi)^2} \mu(k) (L+K_s)\phi(\bm k)\\
&=\int \f{d \bm k}{(2\pi)^2} \mu(k) \Big(\bm k \cdot \nabla_{\bm k} +2+g(k)\Big)\phi(\bm k)\\
&=\int \f{d \bm k}{(2\pi)^2} \Big[\Big(-\bm k \cdot \nabla_{\bm k} +g(k)\Big)\mu(k) \Big]\phi(\bm k)\\
&=\int \f{d \bm k}{(2\pi)^2} \Big[\Big(- k \pa_k +g(k)\Big)\mu(k) \Big]\phi(\bm k).
\label{phi_flow}
\label{phi_flow}
\elgn
\eeqa
In the massless case, the action $S^\Lambda$ is invariant if and only if
\beq
g(k)=\f{k\pa_k \mu(k)}{\mu(k)}.
\eeq
Let $K_\star$ denote the fixed-point disentangler given by the above condition and $S^\Lambda_\star$ the massless action. Then we have
\beq
(L+K_\star)S^\Lambda_\star=0.
\eeq
Note that Eq. \ref{phi_flow} implies that the smeared field $\phi^\Lambda(\bm 0)$ is also invariant under the action of $L+K_\star$.

\begin{figure}[h]
\subfigure[]{\includegraphics[width=\columnwidth]{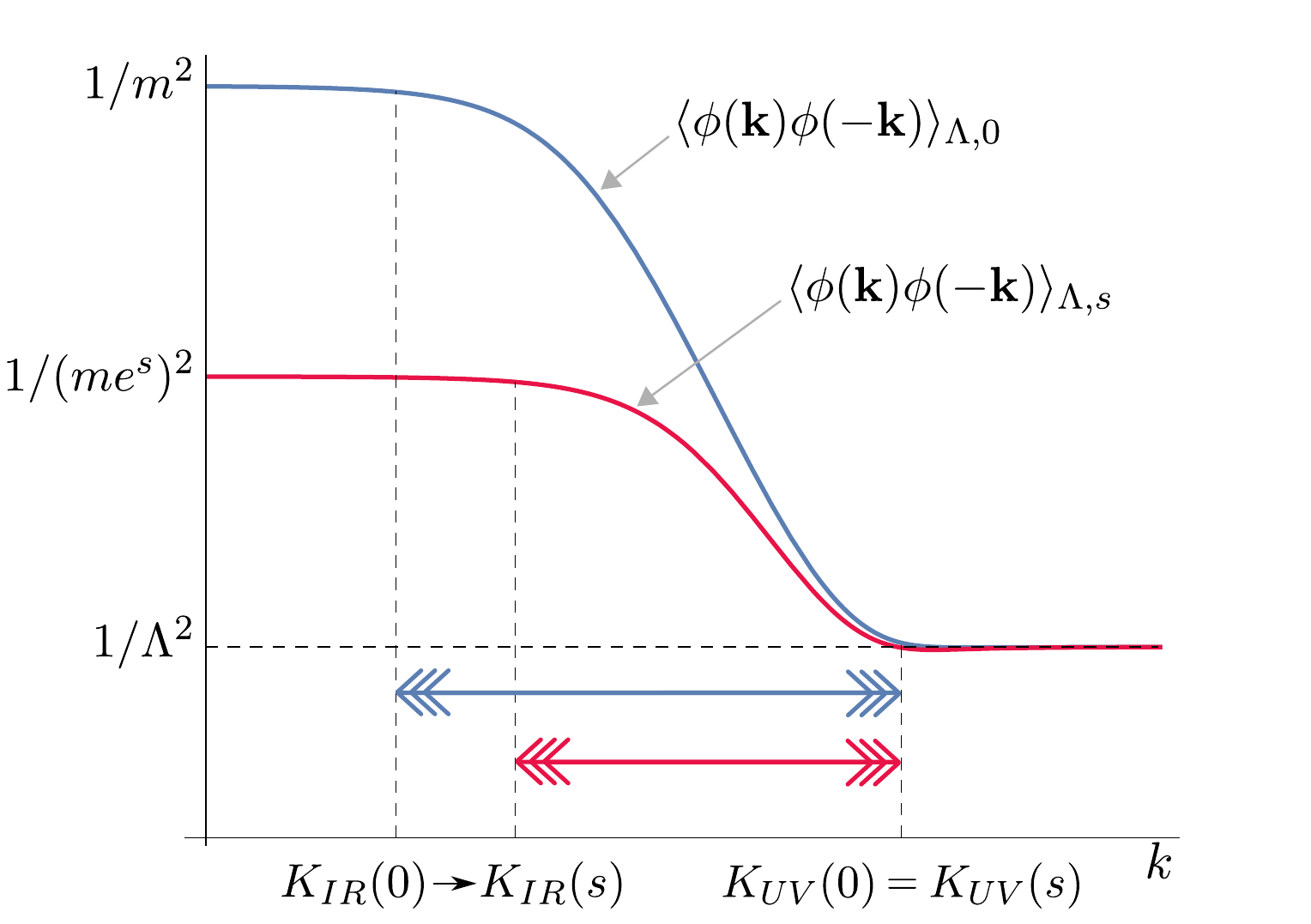}}
\subfigure[]{\includegraphics[width=\columnwidth]{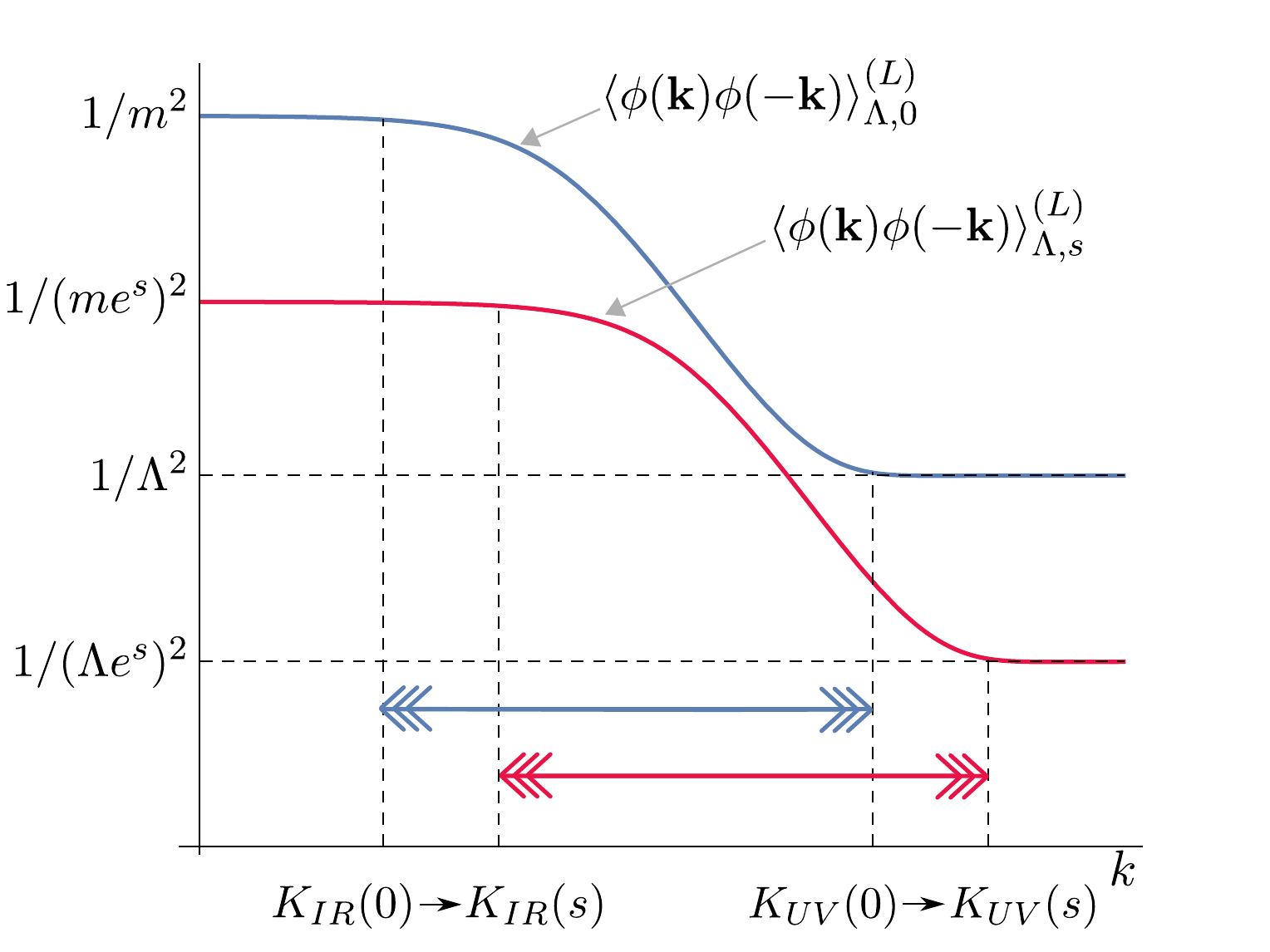}}
\caption{(a) Evolution of $\la \phi(\bm k)\phi(-\bm k) \ra_{\Lambda,s}$ as a function of $s$, generated by $L+K_\star$. The plot is in log-log scale. The correlation is close to a constant outside the window $K_{IR}(s)<k<K_{UV}(s)$, where $K_{IR}\sim me^s$, and $K_{UV} \sim \Lambda$. The width of the window keeps decreasing along the RG flow generated by $L+K_\star$. (b) The evolution of $\la \phi(\bm k)\phi(-\bm k) \ra_{\Lambda,s}^{(L)}$ as a function of $s$, generated by $L$. The plot is in log-log scale. The correlation is close to a constant outside the window $K_{IR}(s)<k<K_{UV}(s)$, where $K_{IR}\sim me^s$, and $K_{UV} \sim \Lambda e^s$. The width of the window remains invariant along the RG flow generated by $L$. }
\label{phiphiflow}
\end{figure}

For a massive theory, $L+K_\star$ generates an RG flow, given by
\beq
(L+K_\star)S^\Lambda[\phi]=\int \f{d\bm k}{(2\pi)^2}  m^2 \mu(k)^2 \phi(\bm k)\phi(-\bm k),
\eeq
or equivalently,
\beqa
\blgn
&S^\Lambda_s[\phi] \equiv e^{s(L+K_\star)}S^\Lambda[\phi]\\
&=\f 1 2\int d\bm x \Big(-\phi^\Lambda(\bm x) \Delta \phi^\Lambda(\bm x)+ m^2 e^{2s}  \phi^\Lambda(\bm x)^2 \Big).
\elgn
\eeqa
We can also compute the correlation function along the RG flow:
\beqa
\blgn
\la \phi(\bm k)\phi(-\bm k) \ra_{\Lambda,s}&\equiv \f{\int [d\phi]e^{-S^\Lambda_s[\phi]}\phi(\bm k)\phi(-\bm k)}{\int [d\phi]e^{-S^\Lambda_s[\phi]}}\\
&=\f{1}{(k^2+m^2 e^{2s})\mu(k)^2}.
\elgn
\eeqa
As \mbox{Fig. \ref{phiphiflow}(a)} shows, the correlation function $\la \phi(\bm k)\phi(-\bm k) \ra_{\Lambda,s}$ approaches a constant $\f{1}{m^2 e^{2s}}$ for small $k$, and a constant $\f{1}{\Lambda^2}$ for large $k$. Therefore, the theory behaves trivially at long distances and short distances. The IR regulator $K_{IR}(s)$ and UV regulator $K_{UV}(s)$ are approximately $m e^s$ and $\Lambda$ respectively. The nontrivial information of the theory for a given value of $s$ is contained in the correlation function $\la \phi(\bm k)\phi(-\bm k) \ra_{\Lambda,s}$ for $K_{IR}(s)<k<K_{UV}(s)$. This window has a decreasing width along the RG flow. This is expected because the disentangler $K_\star$ sequentially removes correlations at different scales. As a comparison, we compute the correlation function for the theory $e^{sL}S^\Lambda[\phi]$. The evolution of the smeared theory generated by $L$ is given by
\beqa
\blgn
&e^{sL}S^\Lambda[\phi]\\
&=\int \f{d\bm k}{(2\pi)^2} (k^2+ m^2) \mu(k)^2 e^{2s}\phi(\bm k e^{s}) e^{2s}\phi(-\bm k e^{s})\\
&=\int \f{d\bm k}{(2\pi)^2} (k^2 + m^2e^{2s}) \mu(k e^{-s})^2 \phi(\bm k ) \phi(-\bm k ).
\elgn
\eeqa
Then we can easily compute the correlation function
\beqa
\blgn
\la \phi(\bm k)\phi(-\bm k) \ra_{\Lambda,s}^{(L)}&\equiv \f{\int [d\phi]e^{-e^{sL}S^\Lambda[\phi]}\phi(\bm k)\phi(-\bm k)}{\int [d\phi]e^{-e^{sL}S^\Lambda[\phi]}}\\
&=\f{1}{(k^2+m^2 e^{2s})\mu(k e^{-s})^2}.
\elgn
\eeqa
As Fig. \ref{phiphiflow}(b) shows, the width of the window containing the nontrivial information of the theory does not change. This is because $L$ only rescales the spacetime and fields, but it does not remove any correlations.

\section{2D free boson CFT}
In this section, we give a brief introduction to the free boson CFT in 2 dimensions. The action is
\beq
S[\phi]=\f 1 2 \int dx_1 dx_2 \Big( (\pa_{x_1} \phi)^2+(\pa_{x_2}\phi)^2   \Big).
\eeq
It is convenient to parametrize the Euclidean plane by complex coordinates:
\beq
z=x_1+ix_2,~~~~\bar z =x_1-ix_2.
\eeq
The primaries in this theory are $\mathbb{1}$, $\pa\phi\equiv \pa_{x_1}\phi-i\pa_{x_2}\phi$, $\bar\pa\phi\equiv \pa_{x_1}\phi+i\pa_{x_2}\phi$ and the vertex operators $\mathcal V_\alpha \equiv :e^{i\alpha \phi}:$. Their conformal dimensions are $(0,0)$, $(1,0)$, $(0,1)$ and $(\f{\alpha^2}{8\pi},\f{\alpha^2}{8\pi})$, respectively.

The correlation of $\pa\phi$ with itself is
\beq
\la \pa\phi(z) \pa\phi(w) \ra=-\f{1}{4\pi} \f{1}{(z-w)^2},
\eeq
from which we can derive the OPE
\beq
 \pa\phi(z) \pa\phi(w) \sim -\f{\mathbb 1}{4\pi} \f{1}{(z-w)^2}.
\eeq
The holomorphic component of the stress tensor $T$ is the regular part of the product of $\pa\phi$ with itself:
\beqa
\blgn
T(z)&=-2\pi :\pa\phi(z)\pa\phi(z):\\
&=-2\pi \lim_{w\to z}\Big( \pa\phi(z)\pa\phi(w)-\la \pa\phi(z)\pa\phi(w)\ra      \Big).
\elgn
\eeqa
Here, the normal ordering $:A(z)B(z):$ of two fields $A(z)$ and $B(w)$ is defined as usual by subtracting all the singular terms of $A(z)B(w)$ in the limit $w\to z$. The OPE of $T(z)$ with $\pa\phi$ can be calculated from Wick's theorem:
\beqa
\blgn
T(z)\pa\phi(w)&=-2\pi :\pa\phi(z)\pa\phi(z):\pa\phi(w)\\
&\sim \f{\pa\phi(z)}{(z-w)^2}\\
&\sim \f{\pa\phi(w)}{(z-w)^2}+\f{\pa^2\phi(w)}{(z-w)}.
\elgn
\eeqa
Furthermore, we can calculate the OPE of $T(z)$ with itself:
\beqa
\blgn
T(z)T(w)&=4\pi^2 :\pa\phi(z)\pa\phi(z)::\pa\phi(w)\pa\phi(w):\\
&\sim \f{1/2}{(z-w)^4}+\f{2T(w)}{(z-w)^2}+\f{\pa T(w)}{(z-w)}.
\elgn
\eeqa
We can read off the central charge $c=1$ from this expression, since the OPE of $T(z)$ with itself for a general CFT is
\beq
T(z)T(w)\sim \f{c/2}{(z-w)^4}+\f{2T(w)}{(z-w)^2}+\f{\pa T(w)}{(z-w)}.
\eeq


\section{Correspondence between sharp and smeared scaling operators}
The massless free boson theory $S[\phi]$ is invariant not only under Euclidean symmetries (translations and rotations), but also under change of scale generated by $L$, and more generally under the conformal group. It has been shown that Euclidean symmetry is preserved in the quasilocal action $S^\Lambda$. However, scale invariance is explicitly broken by the introduction of a UV cutoff, namely the smearing length $\f 1 \Lambda$. Nevertheless, we can define scale invariance with respect to $L+K_\star$. More generally, the smeared theory realizes the whole conformal group although in a quasilocal way, as we explain next.

Since the smearing is diagonal in momentum space, $\phi^\Lambda(\bm k)=\mu(k)\phi(\bm k)$, it changes the integration measure only by a constant factor. Therefore the partition function $Z^\Lambda$ is proportional to the partition function $Z$ of the original CFT. 
\beq
Z^\Lambda =\int [d\phi] e^{-S[\phi^\Lambda]} \propto \int [d\phi^\Lambda] e^{-S[\phi^\Lambda]}=Z.
\eeq
In consequence, we can construct a one-to-one correspondence between smeared fields in the smeared theory and sharp fields in the original CFT. For example, we associate each linear field $\mathcal O(\bm x)$ (linear in terms of $\phi(\bm x)$) in the CFT with a smeared field $\mathcal O^\Lambda(\bm x)$ in the smeared theory by the following relation:
\beq
\mathcal O^\Lambda(\bm x)\equiv \int d\bm y~ \mu(|\bm x -\bm y|)\mathcal O(\bm y) 
\label{smearing}
\eeq
In particular Eq. \ref{smearing} maps the linear local scaling operators $\mathcal O_\alpha(\bm x)$ in the free boson CFT to the linear quasilocal scaling operators $\mathcal O_\alpha^\Lambda(\bm x)$ in the smeared theory.

Local scaling operators are the eigenvectors of $L$ and $R$:
\beqa
\blgn
L~\mathcal O_\alpha(\bm 0)&=-\Delta_\alpha \mathcal O_\alpha(\bm 0),\\
R~\mathcal O_\alpha(\bm 0)&=s_\alpha \mathcal O_\alpha(\bm 0),
\label{local_scaling}
\elgn
\eeqa
while quasilocal scaling operators are the eigenvectors of $L+K_\star$ and $R$:
\beqa
\blgn
(L+K_\star)~\mathcal O_\alpha^\Lambda(\bm 0)&=-\Delta_\alpha \mathcal O_\alpha^\Lambda(\bm 0),\\
R~\mathcal O_\alpha^\Lambda(\bm 0)&=s_\alpha \mathcal O_\alpha^\Lambda(\bm 0).
\label{quasilocal_scaling}
\elgn
\eeqa
The rotation $R$ is unchanged because the smearing function $\mu(x)$ is rotation invariant. Now we associate each linear local scaling operator $\mathcal O_\alpha(\bm 0)$ with a linear smeared operator $\mathcal O_\alpha^\Lambda(\bm 0)\equiv \int d\bm x~ \mu( x)\mathcal O_\alpha(\bm x) $, and show that the equations Eq. \ref{local_scaling} imply the equations Eq. \ref{quasilocal_scaling}.

Since $\mu( x)$ is rotation invariant, obviously $\mathcal O_\alpha^\Lambda(\bm 0)$ is an eigenvector of $R$ with the same conformal spin $s_\alpha$. Applying $(L+K_\star)$ to $\mathcal O_\alpha^\Lambda(\bm 0)$ we get
\beqa
\blgn
&(L+K_\star)~\mathcal O_\alpha^\Lambda(\bm 0)\\
&=\int d\bm x~ \mu( x)(L+K_\star)\mathcal O_\alpha(\bm x) \\
&=\int d\bm x~ \mu( x)\Big( (-\bm x\cdot \nabla_{\bm x}-\Delta_\alpha)\mathcal O_\alpha(\bm x)+\int g(|\bm x-\bm y|)O_\alpha(\bm y) \Big)\\
&=\int d\bm k~ \mu( k) (\bm k\cdot \nabla_{\bm k}+2-\Delta_\alpha+g( k))\mathcal O_\alpha(\bm k)\\
&=\int d\bm k~ (-\bm k\cdot \nabla_{\bm k}-\Delta_\alpha+g( k))\mu( k) \mathcal O_\alpha(\bm k)\\
&=\int d\bm k~ (- k\pa_k+g(k)-\Delta_\alpha)\mu(k) \mathcal O_\alpha(\bm k)\\
&=-\int d\bm k~ \Delta_\alpha \mu(k) \mathcal O_\alpha(\bm k)\\
&=-\Delta_\alpha \mathcal O_\alpha^\Lambda(\bm 0).
\elgn
\eeqa
Therefore, $\mathcal O_\alpha^\Lambda(\bm 0)$ is a smeared scaling operator with the same scaling dimension $\Delta_\alpha$. Thus, for linear operators, we have proved that the smearing process Eq. \ref{smearing} maps local scaling operators in the CFT to the quasilocal scaling operators in the smeared theory. Since the local scaling operators are known for the CFT, we can just use the smearing to find their counterparts in cTNR. For example, the right-moving field $\pa \phi(\bm x)\equiv (\pa_{x_1}-i\pa_{x_2})\phi(\bm x)$ is a primary with scaling dimension $\Delta_{\pa\phi}=1$ and conformal spin $s_{\pa\phi}=1$. Therefore, $\pa\phi^\Lambda(\bm x)$ has the same scaling dimension and conformal spin in the smeared theory. In addition, we can similarly find the quadratic scaling operators such as the holomorphic component of the stress tensor, $T^\Lambda(\bm x)=-2\pi:\pa\phi^\Lambda(\bm x)\pa\phi^\Lambda(\bm x):$ with $\Delta_{T^\Lambda}=s_{T^\Lambda}=2$. Using Wick's theorem, we can then consider any higher powers of the field, and even vertex operators $V^\Lambda_\nu(\bm x)\equiv :e^{i\nu \phi^\Lambda}:$. Furthermore, the operator product expansion (OPE) of the boson CFT is preserved in the smeared theory. For instance, the OPE of the stress tensor $T^\Lambda$ and the primary $\pa\phi^\Lambda(\bm x)$ is the same as in CFT:
\beq
T^\Lambda(z)\pa\phi^\Lambda(w)\sim \f{\pa\phi^\Lambda(w)}{(z-w)^2}+\f{\pa^2\phi^\Lambda(w)}{z-w},
\eeq
where $z$ and $w$ are complex coordinates introduced in the previous section. Finally, the OPE of $T^\Lambda$ with itself gives the value of the central charge $c=1$:
\beq
T^\Lambda(z)T^\Lambda(w)\sim \f{1/2}{(z-w)^4}+\f{2T^\Lambda(w)}{(z-w)^2}+\f{\pa T^\Lambda(w)}{z-w}.
\eeq

Importantly, the quasilocal stress tensor $T^\Lambda(z)$ generates conformal transformations in the smeared theory through the Ward identities as the local stress tensor $T(z)$ does in the CFT. Since $T^\Lambda(z)$ is quasilocal, the conformal symmetries of the smeared theory is realized in a quasilocal fashion.


\begin{thebibliography}{99}
\bibitem{Kstar} We expect a scale-dependent $K_s$ to be generically required in order to generate a proper RG flow. However, in the simple example of a massive free boson the scale-independent $K_{\star}$ suffices.
\bibitem{cTNRanisotropic}
The proposed extension of TNR to the continuum is based on explicitly preserving rotation invariance (in Euclidean spacetime), which requires introducing isotropic smearing of the fields. Alternative extensions of TNR to the continuum are possible using anisotropic smearing. In the limit of strong anisotropy, namely when the smearing affects only the space direction while keeping the fields local in the time direction, we obtain a version of cTNR where it is still straightforward to define a Hilbert space on time slices. There the cMERA appears naturally when using cTNR to coarse-grain an Euclidean path integral with an open (constant time) boundary, exactly as it occurs in the lattice \cite{TNRyieldsMERA}.
\end{thebibliography}
\end{document}